\documentclass[aps,pra,twocolumn,groupedaddress]{revtex4}
\usepackage{graphicx}
\usepackage{multirow}
\usepackage{amsmath}
\usepackage[latin1]{inputenc}
\usepackage{array}
\usepackage{color}
\usepackage[caption=false]{subfig}
\usepackage[french]{babel}
\usepackage{amssymb}
\usepackage{float}
\usepackage{multirow}
\usepackage[normalem]{ulem}

\begin{document}

\title{Limits on Non-Linear Electrodynamics}

\author{M. Fouch\'e}
\affiliation{Institut Non Lin\'eaire de Nice, CNRS and Universit\'e Nice Sophia-Antipolis, 1361 route des Lucioles, 06560 Valbonne, France}

\author{R. Battesti}
\affiliation{Laboratoire
National des Champs Magn\'etiques Intenses (UPR 3228,
CNRS-UPS-UJF-INSA), F-31400 Toulouse Cedex, France, EU}

\author{C. Rizzo}
\email{carlo.rizzo@lncmi.cnrs.fr} \affiliation{Laboratoire
National des Champs Magn\'etiques Intenses (UPR 3228,
CNRS-UPS-UJF-INSA), F-31400 Toulouse Cedex, France, EU}

\date{\today}

\begin{abstract}
In this paper we set a framework in which experiments whose goal is to test QED predictions can be used in a more general
way to test non-linear electrodynamics (NLED) which contains low-energy QED as a special case. We review some of these experiments and we establish limits on the
different free parameters by generalizing QED predictions in the framework of NLED. We finally discuss the implications of these limits on bound systems and isolated charged particles for which QED has been widely and successfully tested.
\end{abstract}


\maketitle

\section{Introduction}\label{Par:intro}

Interactions between electromagnetic fields in vacuum, absent from Maxwell's classical field equations, have been first
predicted in 1934 by Born and Infeld \cite{Born_1934} in the framework of a new field theory. The main goal of this intrinsically non linear theory was to solve the difficulty related to the fact that the self energy of a point charge is infinite by assuming the existence of an {\it absolute field} \cite{Born_1934} in nature. Born and Infeld have chosen the absolute field amplitude as the amplitude of the electric field created by an electron at a distance equivalent to its classical radius, in others words by equating the classical self-energy of the electron with its mass energy at rest.

In the following years (1935 and 1936), Euler and Kockel \cite{Euler_1935} and then Heisenberg and Euler \cite{Heisenberg_1936} established their own non linear electromagnetic theory, based on the Dirac's vacuum model \cite{Dirac_1934}. The related effective Lagrangian has been validated in 1951 by Schwinger \cite{Schwinger_1951} in the framework of Quantum ElectroDynamics (QED) field theory, and it is nowadays accepted as the mathematical description of field interactions.

Born-Infeld and Heisenberg-Euler theories are two different forms of what is called Non Linear ElectroDynamics (NLED). NLED is a general framework of theories all
describing field-field interactions and predicting a large panel of phenomena going from variations of light velocity in vacuum in the presence of electromagnetic fields to photon-photon scattering but also changes in the long range electromagnetic potential induced by charged particles,
as discussed in this paper.

QED is considered as a very well tested theory. It is indisputable that some of QED numerical predictions has been experimentally verified with an astonishing precision (see e.g. reference \cite{Hanneke_2008}). Thus, it is legitimate to wonder whether
alternative NLED forms have been definitively ruled out or not. Moreover, in the framework of QED itself, it is worthwhile to understand what is the
impact of QED tests for bound or isolated particles into the photon sector, where tests are hardly found. In other words, have complex experiments looking for photon-photon interactions still an impact on QED or can they be considered as a somewhat useless technological prowess whose results are known in advance ?

In this paper we set a framework in which experiments whose goal is to test QED predictions can be used in a more general way to test different NLED theories, which contain low-energy QED \footnote{Low-energy QED is valid for fields varying on scales large compared to the electron Compton wavelength.} as a special case. This can be done by properly parametrizing effective lagrangians.
Actually, assuming that Lorentz invariance holds in vacuum, the mathematical description of all forms of Lorentz-invariant NLED, also known as NLED theory of Pleba\'nski class\,\cite{Schellstede_2015,Boillat_1970,Plebanski_1970}, are given
by a lagrangian depending only on the two Lorentz-invariants $\mathcal{F}$ and $\mathcal{G}$:
\begin{eqnarray}
\mathcal{F} &=& \epsilon_{0}E^2 - {B^2 \over \mu_{0}}, \label{Eq:F}\\
\mathcal{G} &=&\sqrt{\epsilon_{0} \over \mu_{0}} \textbf{E} \cdot \textbf{B},
\label{Eq:G}
\end{eqnarray}
with $\textbf{E}$ and $\textbf{B}$ the electric and magnetic fields, $\epsilon_0$ the vacuum permittivity and $\mu_0$
the vacuum permeability. For weak electromagnetic fields, the lagrangian can be written as a power expansion of $\mathcal{F}$ and $\mathcal{G}$\,\cite{Battesti_2013}:
\begin{equation}\label{Eq:L_genform}
\mathcal{L} = \sum_{i=0}^\infty \sum_{j=0}^\infty c_{i,j} \mathcal{F}^i
\mathcal{G}^{j}.
\end{equation}
The number of free parameters $c_{i,j}$ is infinite, but it is generally accepted that the
lowest orders in the fields are sufficient to describe the phenomena induced in most experiments. The Lagrangian becomes:
\begin{eqnarray}
\mathcal{L} &=& \mathcal{L}_0 + \mathcal{L}_{\mathrm{NL}} \label{Eq:L}\\
\mathrm{with}\quad \mathcal{L}_0 &=& \frac{1}{2}\mathcal{F} \label{Eq:L_0}\\
\mathrm{and}\quad \mathcal{L}_{\mathrm{NL}} &\simeq& c_{0,1}\mathcal{G} + c_{2,0}\mathcal{F}^2 + c_{0,2}\mathcal{G}^2 + c_{1,1}\mathcal{F}\mathcal{G}.
\label{Eq:L_NL}
\end{eqnarray}
The lowest order term $\mathcal{L}_0$ gives the classical Maxwell lagrangian, with $c_{1,0} = 1/2$. The non-linear correction $\mathcal{L}_{\mathrm{NL}}$
depends on four parameters: $c_{0,1}$, $c_{2,0}$, $c_{0,2}$ and $c_{1,1}$.

To describe the non-linear response of vacuum, we treat it as a polarizable medium. One can use the Maxwell equations together with the constitutive equations
related to the lagrangian as follows \cite{LandauLifshitz}:

\begin{eqnarray}
\textbf{P} &=& \frac{\partial\mathcal{L}}{\partial\textbf{E}} - \epsilon_0 \textbf{E}, \label{Eq:D}\\
\textbf{M} &=& \frac{\partial\mathcal{L}}{\partial\textbf{B}} - \frac{\textbf{B}}{\mu_0}\label{Eq:H}.
\end{eqnarray}
$\textbf{P}$ is the polarization and $\textbf{M}$ is the magnetization. Using equations\,(\ref{Eq:L}), (\ref{Eq:L_0}) and (\ref{Eq:L_NL}), one obtains:
\begin{eqnarray}
\textbf{P} &=& c_{0,1}\sqrt{\frac{\epsilon_0}{\mu_0}}\textbf{B} \nonumber\\
&+& 4c_{2,0}\epsilon_0\mathcal{F}\textbf{E} \nonumber\\
&+& 2c_{0,2}\sqrt{\frac{\epsilon_0}{\mu_0}}\mathcal{G}\textbf{B} \nonumber\\
&+& c_{1,1}\Big(2\epsilon_0\mathcal{G}\textbf{E} +
\sqrt{\frac{\epsilon_0}{\mu_0}}\mathcal{F}\textbf{B}\Big), \label{Eq:P}\\
\textbf{M} &=& c_{0,1}\sqrt{\frac{\epsilon_0}{\mu_0}}\textbf{E} \nonumber\\
&-& 4c_{2,0}\mathcal{F}\frac{\textbf{B}}{\mu_0} \nonumber\\
&+& 2c_{0,2}\sqrt{\frac{\epsilon_0}{\mu_0}}\mathcal{G}\textbf{E} \nonumber\\
&-& c_{1,1}\Big(2 \mathcal{G} \frac{\textbf{B}}{\mu_0} - \sqrt{\frac{\epsilon_0}{\mu_0}}\mathcal{F}\textbf{E}\Big). \label{Eq:M}
\end{eqnarray}
Starting from these constitutive equations, one can study the phenomenology associated with the four parameters
$c_{0,1}$, $c_{2,0}$, $c_{0,2}$ and $c_{1,1}$. Corresponding experiments are then able to discriminate different forms of
non linear electrodynamics.

The scope of our work is not to provide a review on theoretical activities and experimental proposals on NLED. Our main goal is to use some existing experimental results to set limits on NLED in a unified framework. In particular, we aim to give a unified approach to compare the results on light propagation in vacuum and experiments on bound systems and isolated particles.

In the following we first give some examples of NLED lagrangians, in particular the Heisenberg and Euler lagrangian predicted in the framework of QED. Then, experimental constraints on the
$c_{i,j}$ parameters are reviewed. We start with photon-photon interaction experiments. Discussing vacuum magnetic birefringence
and photon-photon scattering, we show that a limit on vacuum magnetic birefringence cannot directly give a limit
on the photon-photon scattering cross section as claimed in several papers\,\cite{Bregant_2008, Zavattini_2012, DellaValle_2014}. We finally discuss the implications of this type of lagrangian on bound systems and isolated charged particles for which QED has been widely and successfully tested.


\section{Some effective non linear lagrangians}\label{Par:Lagrangian}

To illustrate the general form of the non-linear lagrangian given in equation\,(\ref{Eq:L_NL}), we focus on some of the most well-known ones.

\subsection{Heisenberg and Euler effective lagrangian}

The generally accepted effective lagrangian is the one established in 1936 by Heisenberg and Euler \cite{Heisenberg_1936} in the
framework of QED. It generalized at all orders the previous work of Euler and Kockel in 1935 \cite{Euler_1935}. Vacuum is assumed to
be C, P and T invariant. This implies that the coefficients $c_{i,j}$ with
an odd index $j$ are null, in particular $c_{0,1}=0$ and $c_{1,1}=0$. The non-linear correction of the lagrangian is then:
\begin{equation}\label{Eq:L_HE}
\mathcal{L}_{\mathrm{NL}} = c_{2,0}\mathcal{F}^2 + c_{0,2}\mathcal{G}^2.
\end{equation}
Following the Euler and Kockel result \cite{Euler_1935}, the value of $c_{2,0}$ and $c_{0,2}$ can be written as:
\begin{eqnarray}
c_{2,0} &=& {2\alpha^2 \hbar^3 \over 45 m_{e}^4 c^5}\label{Eq:c20_HE}\\
&=& {\alpha \over
90 \pi} {1 \over \epsilon_0 E_\mathrm{cr}^2} = {\alpha \over 90 \pi}
{\mu_0 \over B_\mathrm{cr}^2}\\
&\simeq& 1.66 \times 10^{-30}~\left[{m^3
\over J}\right],\label{Eq:c20_HE_number}\\
c_{0,2} &=& 7 c_{2,0}\label{Eq:c02_HE_number},
\end{eqnarray}
and therefore
\begin{equation}
\mathcal{L}_{\mathrm{NL}} = {\alpha \over
90 \pi} {1 \over \epsilon_0 E_{cr}^2}[
\mathcal{F}^2+7\mathcal{G}^2]. \label{LEK}
\end{equation}
where $\alpha = e^2 / 4\pi\epsilon_0\hbar c$ is the fine
structure constant, $e$ the elementary charge, $\hbar$ the Planck
constant $h$ divided by $2\pi$. $E_\mathrm{cr} = m_e^2c^3 / e\hbar$
is a quantity obtained by combining the fundamental constant
$m_e$, the electron mass, $c$, $e$ and $\hbar$. It has
the dimensions of an electric field, and it is called the critical
electric field. Its value is $E_\mathrm{cr} = 1.3 \times 10^{18}$ V/m. A
critical magnetic field can also be defined in the same manner:
$B_\mathrm{cr} = E_\mathrm{cr} / c = m_e^2c^2 / e\hbar =
4.4 \times 10^9$\,T.

The existence of several phenomena can be predicted using this lagrangian, as detailed in reference\,\cite{Battesti_2013}. As long as QED is
supposed to be correct in the presently accepted form, the value of the $c_{i,j}$ coefficients are fixed. Therefore, no prediction
contains any free parameter. The values of the physical quantities to be measured simply correspond to linear combinations of powers
of the fundamental constants $\alpha$, $\hbar$, $m_e$ and $c$.

\subsection{Born-Infeld effective lagrangian}

The Born-Infeld effective lagrangian \cite{Born_1934} is a well known example of NLED theory developed in 1934, even before the
Heisenberg-Euler one. It was introduced to remove the problem of classical self energy of elementary particles which is infinite.
The lagrangian is established from the postulate that there exists an ``absolute field'' $E_\mathrm{abs}$ corresponding to the upper
limit of a purely electric field. The lagrangian is:
\begin{equation}
\mathcal{L} = \epsilon_0 E_\mathrm{abs}^2 \left(-\sqrt{1-\frac{\mathcal{F}}{\epsilon_0 E_\mathrm{abs}^2}-\frac{\mathcal{G}^2}{(\epsilon_0 E_\mathrm{abs}^2)^2}}+1\right). \label{Eq:LBI}
\end{equation}
$E_\mathrm{abs}$ is a free parameter corresponding to a new fundamental constant to be determined.
If we assume that $\left(\frac{\mathcal{F}}{\epsilon_0 E_\mathrm{abs}^2}-\frac{\mathcal{G}^2}{\epsilon_0 E_\mathrm{abs}^4}\right) \ll 1$,
the lagrangian can be developed and, at the lowest orders in the fields, it can be written as:
\begin{equation}
\mathcal{L} \simeq \frac{1}{2}\mathcal{F} +\frac{1}{8\epsilon_0 E_\mathrm{abs}^2}\mathcal{F}^2 + \frac{1}{2 \epsilon_0 E_\mathrm{abs}^2}\mathcal{G}^2. \label{LBIap}
\end{equation}
The corresponding $c_{i,j}$ parameters are:
\begin{eqnarray}
c_{1,0} &=& \frac{1}{2},\\
c_{0,1} &=& c_{1,1} = 0,\\
c_{2,0} &=& \frac{1}{8 \epsilon_0 E_\mathrm{abs}^2},\\
c_{0,2} &=& \frac{1}{2 \epsilon_0 E_\mathrm{abs}^2} = 4c_{2,0}.
\end{eqnarray}
Comparing these terms with the ones obtained in equations\,(\ref{Eq:c20_HE}) and (\ref{Eq:c02_HE_number}) with the Heisenberg-Euler lagrangian, one can see that no value of $E_\mathrm{abs}$ allows the parameters to coincide. Both lagrangians are essentially different and will lead to different non-linear properties. Experimental tests are thus crucial to establish which one is valid. Some examples of possible experiments will be present in the following section, but other configurations can be found for instance in references\,\cite{Denisov_2000, Davila_2014}.

The absolute field constant was estimated in reference\,\cite{Born_1934}. It was related to the ``radius'' of the
electron $r_0$ as follows: $E_\mathrm{abs} =
e/4 \pi \epsilon_0
r_0^2$.
Using the classical electron radius $r_0 = e^2/4 \pi \epsilon_0m_\mathrm{e} c^2$, one finds $E_\mathrm{abs} \simeq 2\times 10^{20}$\,V/m, which corresponds to a $c_{2,0}$ about four times smaller than the one of Heisenberg and Euler.

\begin{figure}[h]
\begin{center}
\includegraphics[width=8cm]{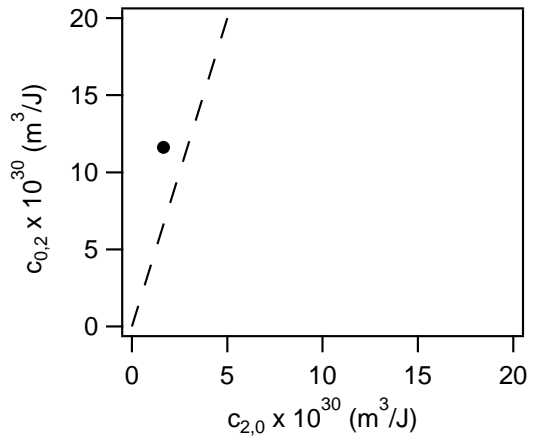}
\caption{\label{Fig:c02_c20_HE_BI} Born-Infeld prediction and Heisenberg-Euler prediction in the $(c_{2,0},c_{0,2})$ parameter space.
The Born-Infled prediction is represented by a straight line, while the Heisenberg-Euler one is a point.}
\end{center}
\end{figure}

Let's recall that the Born and Infeld choice of the absolute field is arbitrarily related to the pointlike particle known at their epoch, the electron.
The absolute field is therefore a free parameter of the Born-Infeld theory that can be experimentally
constrained or measured. The ratio between $c_{2,0}$ and $c_{0,2}$ is however fixed. In the $(c_{2,0},c_{0,2})$ parameter space,
Born-Infeld prediction is thus represented by a straight line, while the Heisenberg-Euler one is represented by a point, as shown in Fig.\,\ref{Fig:c02_c20_HE_BI}.

\subsection{Lagrangian in the string theory framework}

Both Heisenberg-Euler and Born-Infeld lagrangians at the lowest orders in the fields can be considered as special cases
of a more general one obtained in the framework of string theory \cite{Gibbons_2001} which gives a more general interest to
 the field of NLED. This is discussed in details in reference\,\cite{Abalos_2015}. This lagrangian can be written as:
\begin{equation}
\mathcal{L} = \frac{1}{2}\mathcal{F} +\frac{\gamma}{4} \left[(1-b)\mathcal{F}^2 + 6 \mathcal{G}^2\right] \label{LST}
\end{equation}
where $\gamma$ and $b$ are two free parameters. The corresponding $c_{i,j}$ parameters are:

\begin{eqnarray}
c_{1,0} &=& \frac{1}{2},\\
c_{0,1} &=& c_{1,1} = 0,\\
c_{2,0} &=& \frac{\gamma}{4}(1-b),\\
c_{0,2} &=& \frac{3}{2}\gamma.
\end{eqnarray}
The Born-Infeld Langrangian is recovered with $b=-1/2$ and $\gamma = 1/3\epsilon_0 E_\mathrm{abs}^2$. For the Heisenberg-Euler
prediction, one has $b=1/7$ and $\gamma = 7\alpha/135\pi\epsilon_0 E_\mathrm{cr}^2$.

\section{Light propagation in vacuum}\label{Par:Light}

The expected non linear optical phenomena in vacuum are reviewed in Ref.\,\cite{Battesti_2013}. It goes from birefringence effects
induced by electric or magnetic fields, to vacuum dichro\"{\i}sm, photon splitting, photon-photon scattering and second harmonic generation.
In the following, we will focus on the two non linear effects whose experimental observation has been sought quite recently: the magnetic birefringence and
photon-photon scattering.

\subsection{Magnetic Birefringence}

Birefringence can be induced by an electric field, a magnetic field or a combination of both. However, experiments are mostly devoted
to magnetically induced effects. This is due to the fact that the same level of effect is obtained in the presence of a $B$ field or
an electric field $E$ equal to $cB$. From a technological point of view, magnetic fields of several tesla are easier to produce than
electric fields of about 1\,GV\,m$^{-1}$.

\subsubsection{Expected birefringence}

The calculation of the birefringence induced by a transverse static magnetic field, using the general lagrangian given by
equations\,(\ref{Eq:L}) to (\ref{Eq:L_NL}), can be found in reference\,\cite{Pinto_2006}. In the following, we only briefly give the main steps.

The total magnetic field corresponds to the sum of the static magnetic field $\textbf{B}_0$ and the one of the propagating wave
$\textbf{B}_\omega$: $\textbf{B} = \textbf{B}_\omega + \textbf{B}_0$. The electric field associated to the propagating wave is
$\textbf{E}_\omega$. Introducing these quantities in equations\,(\ref{Eq:P}) and (\ref{Eq:M}) and keeping only the $\omega$ component, we obtain:
\begin{eqnarray}
\textbf{P}_\omega &=& - \frac{4 \epsilon_0 c_{2,0}}{\mu_0}B_0^2\textbf{E}_\omega \nonumber\\
&+& \frac{2 \epsilon_0 c_{0,2}}{\mu_0}(\textbf{E}_\omega\cdot\textbf{B}_0)\textbf{B}_0 \nonumber\\
&+& \sqrt{\frac{\epsilon_0}{\mu_0}}\left(c_{0,1}-\frac{c_{1,1}}{\mu_0} B_0^2\right)\textbf{B}_\omega \nonumber\\
&-& \sqrt{\frac{\epsilon_0}{\mu_0}}\frac{2 c_{1,1}}{\mu_0}(\textbf{B}_\omega\cdot\textbf{B}_0)\textbf{B}_0, \label{Eq:D_omega}\\
\textbf{M}_\omega &=& \frac{4 c_{2,0}}{\mu_0^2}B_0^2\textbf{B}_\omega \nonumber\\
&+& \frac{8 c_{2,0}}{\mu_0^2}(\textbf{B}_\omega\cdot\textbf{B}_0)\textbf{B}_0 \nonumber\\
&-& \sqrt{\frac{\epsilon_0}{\mu_0}}\left(-c_{0,1}+\frac{c_{1,1}}{\mu_0} B_0^2\right)\textbf{E}_\omega\nonumber\\
&-& \sqrt{\frac{\epsilon_0}{\mu_0}}\frac{2 c_{1,1}}{\mu_0}(\textbf{E}_\omega\cdot\textbf{B}_0)\textbf{B}_0.
\label{Eq:H_omega}
\end{eqnarray}

We define the static magnetic field direction as the x-direction. This magnetic field is transverse to the light propagation,
supposed to be along the z-direction. We assume the existence of plane wave eigenmodes with refractive index $n$:
\begin{equation}
\textbf{E}_\omega(\textbf{r},t) = \textbf{E}_{0}\mathrm{e}^{i\omega\left(\frac{n}{c} \textbf{e}_z\cdot \textbf{r} - t\right)}.
\end{equation}
Injected into the Maxwell equations, one gets in the polarization plane $(x,y)$:
\begin{widetext}
\[
\begin{pmatrix}
   n^2\left(\frac{4c_{2,0}}{\mu_0}B_0^2 - 1\right) + 2 + \frac{2(c_{0,2} - 2c_{2,0})}{\mu_0}B_0^2 & \frac{2nc_{1,1}}{\mu_0}B_0^2 \\
   \frac{2nc_{1,1}}{\mu_0}B_0^2 & n^2\left(\frac{12c_{2,0}}{\mu_0}B_0^2 - 1\right) + 2 - \frac{4c_{2,0}}{\mu_0}B_0^2
\end{pmatrix}
\textbf{E}_\omega = \textbf{E}_\omega.
\]
\end{widetext}

We can first note that the $c_{0,1}$ term has canceled out and thus does not contribute to the propagation of light. The diagonal terms
correspond to the Cotton-Mouton effect. In this case, the eigenmodes are parallel and perpendicular to the magnetic field.
The corresponding index of refraction are:
\begin{eqnarray}
n_{\|} = 1 + \frac{c_{0,2}}{\mu_{0}}B_{0}^2,\\
n_{\bot} = 1 + \frac{4c_{2,0}}{\mu_{0}}B_{0}^2,
\end{eqnarray}
where $n_{\|}$ is the index of refraction for light polarized parallel to the external magnetic field and $n_{\bot}$ is the index
of refraction for light polarized perpendicular to the external magnetic field. While the refractive index $n_{\|}$ depends only
on $c_{0,2}$, $n_{\bot}$ depends only on $c_{2,0}$. Since dispersive effects can be neglected, $n_{\|}$ and $n_{\bot}$ have to be always greater than 1 and $c_{0,2}$ and $c_{2,0}$ have to be greater than 0.

The anisotropy $\Delta n$ is equal to:
\begin{equation}\label{Deltan}
\Delta n_\mathrm{CM} = n_{\|} - n_{\bot} = \frac{c_{0,2}-4c_{2,0}}{\mu_{0}}B_{0}^2,
\end{equation}
and depends on both parameters. Let's note that in the case of Heisenberg-Euler lagrangian, one gets:
\begin{equation}
\Delta n_\mathrm{CM, HE} = \frac{3c_{2,0}}{\mu_0}B_0^2 = \frac{2\alpha^2\hbar^3}{15 \mu_0 m_e^4 c^5}B_0^2. \label{Eq:CM_HE}
\end{equation}
On the other hand, with the Born-Infeld lagrangian, no Cotton-Mouton effect is expected \cite{Boillat_1970,Plebanski_1970,Bialynicki_1983} since we get:
\begin{equation}
\Delta n_\mathrm{CM,BI} = 0.
\end{equation}

The non-diagonal terms can be interpreted as a magnetic Jones birefringence, with a linear birefringence along axis which are
at $\pm 45\char23$ relative to the direction of the static magnetic field. The corresponding difference of refractive index is \cite{Note_Jones}:
\begin{equation}
\Delta n_\mathrm{J} = n_{+ 45\char23} - n_{- 45\char23} =  \frac{2c_{1,1}}{\mu_{0}}B_{0}^2.
\end{equation}

\subsubsection{Experimental limits}

Two types of experiments have been realized to measure this variation of the light velocity in the presence of a transverse
magnetic field \cite{Battesti_2013}. The first one is based on interferometers with separated arms, such as the Michelson-Morley
interferometer. The basic idea is to look at the interference displacement when a magnetic field is applied on one of the arm.
This type of configuration has the advantage to directly measure one of the parameters, $c_{0,2}$ or $c_{2,0}$ if the magnetic
field is oriented parallel or perpendicular to the light polarization.

In 1940, Farr and Banwell reported results obtained using an interferometer where one of the two arms is immersed in a 2\,T magnetic field. The
measured relative variation of light velocity was less than $2\times 10^{-9}$\,\cite{Banwell_1940}. The light polarization with respect to the magnetic field was not clearly stated. For the sake of argument, assuming that one can infer limits on the $c_{i,j}$ parameters from their measurements, we obtain:
\begin{eqnarray}
c_{2,0}&<&1.6\times10^{-16}~\mathrm{m}^3\mathrm{J}^{-1},\\
c_{0,2}&<&6.3\times10^{-16}~\mathrm{m}^3\mathrm{J}^{-1},\\
c_{1,1}&<&6.3\times10^{-16}~\mathrm{m}^3\mathrm{J}^{-1}.
\end{eqnarray}
Anyway, these limits are at 14 orders of magnitude from the QED predictions (see equations,(\ref{Eq:c02_HE_number}) and (\ref{Eq:c20_HE_number}).

The second type of experiments is based on polarimetry. The principle is to measure the magnetic birefringence via the ellipticity induced
on a linearly polarized laser beam propagating in a transverse magnetic field \cite{Iacopini_1979}. In this case, one measures the
difference of refractive index and not directly the refractive index. Therefore, concerning the Cotton-Mouton configuration,
the measurement cannot by itself constrain both $c_{0,2}$ and $c_{2,0}$ but only a particular linear combination of the two free
parameters: $c_{0,2} -4 c_{2,0}$. Let's note finally that, even if one measures the value predicted by the Heisenberg and Euler
lagrangian for $\Delta n_\mathrm{CM}$ i.e. $3 c_{2,0}
^\mathrm{HE}
B_0^2/\mu_0$, this cannot be considered in principle the definitive demonstration
that this lagrangian is correct. Any lagrangian with $c_{0,2} - 4c_{2,0} = 3 c_{2,0}^\mathrm{HE}$ predicts the same value.

\begin{figure}[h]
\begin{center}
\includegraphics[width=8cm]{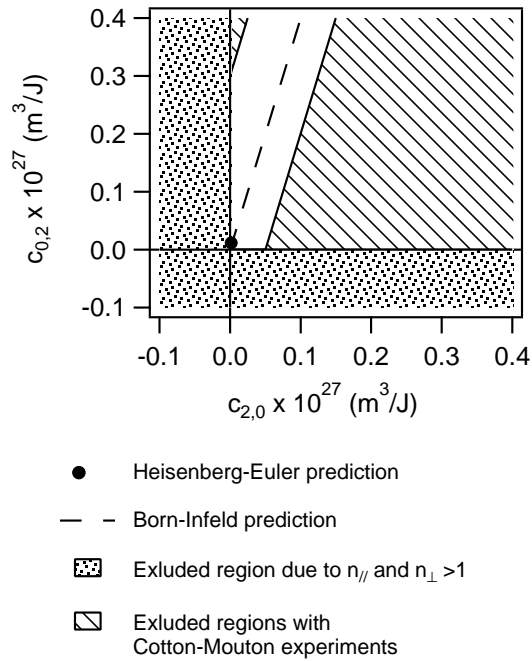}
\caption{\label{Fig:c02_c20_Exp_limits} Best experimental limits on $c_{0,2}$ and $c_{2,0}$ parameters. Striped areas: excluded region obtained with the result of reference\,\cite{DellaValle_2014}. Point: Heisenberg-Euler prediction.
Dashed line: Born-Infeld prediction. The point seems superimposed with the dashed line due to the scale. Dotted areas: excluded regions due to the fact that $n_{\|}$ and $n_\bot > 1$.}
\end{center}
\end{figure}

The most advanced experiments in this domain are the one of the PVLAS collaboration\,\cite{DellaValle_2014} and the one of the BMV
group\,\cite{Cadene_2014}. The direction of the static magnetic field is at $45\char23$ compared to the light polarization,
corresponding to the Cotton-Mouton configuration. Experiments measure $\Delta n_{\mathrm{CM}}$ with an error $\delta \Delta n_{\mathrm{CM}}$.
This corresponds in the $(c_{0,2},c_{2,0})$ parameter plane to two regions of exclusion:
\begin{eqnarray}
c_{0,2} &<& 4c_{2,0} +\mu_0(\Delta n_{\mathrm{CM}} + \delta \Delta n_{\mathrm{CM}}),\\
c_{0,2} &>& 4c_{2,0} +\mu_0(\Delta n_{\mathrm{CM}} - \delta \Delta n_{\mathrm{CM}}).
\end{eqnarray}
The best limit is given in Ref.\,\cite{DellaValle_2014} with $\Delta n = (0.4\pm 2.0)\times 10^{-22} B_0^2$ at $1\sigma$, corresponding to:
\begin{eqnarray}
c_{0,2} &<& 4c_{2,0} +3\times10^{-28} ~\mathrm{m}^3\mathrm{J}^{-1},\\
c_{0,2} &>& 4c_{2,0} -2 \times10^{-28}~\mathrm{m}^3\mathrm{J}^{-1}.
\end{eqnarray}
These limits are summarized in Fig.\,\ref{Fig:c02_c20_Exp_limits}.

Finally, to give a limit one the $c_{1,1}$ parameter, one should use the Jones configuration, with the light polarization parallel or perpendicular to the magnetic field
as discussed in references \cite{Pinto_2006} and \cite{Millo_2009}. In the last reference, Millo and Faccioli have also estimated the magnitude of this effect within the standard model
using Quantum ChromoDynamics chiral perturbation theory obtaining that $c_{1,1}$ is expected to be at least 20 orders of magnitude smaller than $c_{2,0}^\mathrm{HE}$. Anyway, no one has ever done such a measurement.

\subsection{Photon-photon scattering}

Testing low-energy QED with ultra-intense lasers is widely discussed in the literature, with in particular the direct observation of photon-photon scattering. Recent reviews can be found in references\,\cite{Bulanov_2011, DiPiazza_2012, King_2012, Narozhny_2015, King_2016}. In the following, we will focus on the experiment which has reported the best experimental limit up to now\,\cite{Bernard_2000}.

The most simple experiment to look at photon-photon scattering in vacuum consists in two colliding laser beams as proposed in reference\,\cite{Moulin_1996}. The calculation of the corresponding total photon-photon scattering cross section for unpolarized light with the Heisenberg-Euler or the Born-Infeld lagrangian can be found for example in reference\,\cite{Davila_2014}. The number of
scattered photons can be enhanced by using a third beam which stimulates the reaction\,\cite{Varfolomeev_1966}.
In this configuration, the link between the $c_{i,j}$ coefficients and the measurement of the number of scattered photons
can be established following the approach proposed in references\,\cite{Moulin_1999} and \cite{Bernard_2000} where a third-order
non linear effective susceptibility $\chi_v^3$ is introduced, as in classical nonlinear optics. Here, we only present the main steps
of the calculations.

In elastic scattering, the energy and momentum conservation holds, corresponding to:
\begin{eqnarray}
\textbf{k}_4 = \textbf{k}_1+\textbf{k}_2-\textbf{k}_3,\\
\omega_4 = \omega_1+\omega_2
-\omega_3,
\end{eqnarray}
with $\textbf{k}_i$ the wave vector of laser beam number $i$ and $\omega_i$ its frequency multiply by $2\pi$.
The three incoming beams are 1, 2 and 3, while beam number 4 is the scattered one. Using equations\,(\ref{Eq:P})
and (\ref{Eq:M}) and keeping only the $\omega_4$ component, we obtain:
\begin{eqnarray}
\textbf{P}_{\omega_4} &=& \epsilon_0^2 E_1E_2\overline{E}_3 \bigg[ 2c_{2,0}\textbf{K}_{P_{20}} + \frac{c_{02}}{2}\textbf{K}_{P_{02}}\nonumber\\
&+& \frac{c_{11}}{2}\left( \textbf{K}_{P_{11,1}} + \textbf{K}_{P_{11,2}} \right)\bigg], \label{Eq:P_omega4}\\
&=& \epsilon_0^2 E_1E_2\overline{E}_3 \textbf{K}_P, \label{Eq:KP}\\
\textbf{M}_{\omega_4} &=& c\epsilon_0^2 E_1E_2\overline{E}_3 \bigg[ -2c_{2,0}\textbf{K}_{P_{11,2}} + \frac{c_{02}}{2}\textbf{K}_{P_{11,1}}\nonumber\\
&-& \frac{c_{11}}{2}\left( -\textbf{K}_{P_{02}} + \textbf{K}_{P_{20}} \right)\bigg]\label{Eq:M_omega4},\\
&=& c\epsilon_0^2 E_1E_2\overline{E}_3 \textbf{K}_M, \label{Eq:KM}
\end{eqnarray}
where $\textbf{E}_i$ is the electric field of beam number $i$. The geometrical factors are:
\begin{eqnarray}
\textbf{K}_{P_{20}}&=&{\bf{u_1}}\left({\bf{u_2}}.{\bf{u_3}} - {\bf{v_2}}.{\bf{v_3}}\right)\nonumber\\
&+& {\bf{u_2}}\left({\bf{u_1}}.{\bf{u_3}} - {\bf{v_1}}.{\bf{v_3}}\right)\nonumber\\
&+& {\bf{u_3}}\left({\bf{u_1}}.{\bf{u_2}} - {\bf{v_1}}.{\bf{v_2}}\right),\\
\textbf{K}_{P_{02}}&=&{\bf{v_1}}\left({\bf{u_2}}.{\bf{v_3}} + {\bf{v_2}}.{\bf{u_3}}\right)\nonumber\\
&+& {\bf{v_2}}\left({\bf{u_1}}.{\bf{v_3}} + {\bf{v_1}}.{\bf{u_3}}\right)\nonumber\\
&+& {\bf{v_3}}\left({\bf{u_1}}.{\bf{v_2}} + {\bf{v_1}}.{\bf{u_2}}\right),\\
\textbf{K}_{P_{11,1}}&=&{\bf{u_1}}\left({\bf{u_2}}.{\bf{v_3}} + {\bf{v_2}}.{\bf{u_3}}\right)\nonumber\\
&+& {\bf{u_2}}\left({\bf{u_1}}.{\bf{v_3}} + {\bf{v_1}}.{\bf{u_3}}\right)\nonumber\\
&+& {\bf{u_3}}\left({\bf{
u_1}}.{\bf{v_2}} + {\bf{v_1}}.{\bf{u_2}}\right),\\
\textbf{K}_{P_{11,2}}&=&{\bf{v_1}}\left({\bf{u_2}}.{\bf{u_3}} - {\bf{v_2}}.{\bf{v_3}}\right)\nonumber\\
&+& {\bf{v_2}}\left({\bf{u_1}}.{\bf{u_3}} - {\bf{v_1}}.{\bf{v_3}}\right)\nonumber\\
&+& {\bf{v_3}}\left({\bf{u_1}}.{\bf{u_2}} - {\bf{v_1}}.{\bf{v_2}}\right)
\end{eqnarray}
The unit vectors $\bf{u_i}$ and $\bf{v_i}$ indicate the direction of the electric field (i.e. the polarization)
of the photon beam $i$ and the direction of the corresponding magnetic field. The geometrical factors depend on
the directions of the incident beam and on their polarizations.

The propagation equation for the electric field $\textbf{E}_4$ is obtained thanks to Maxwell's equations in the slow
varying wave approximation \cite{Moulin_1999,Bernard_2000}:
\begin{eqnarray}
&&\nabla^2 \textbf{E}_4 - \frac{1}{c^2}\frac{\partial^2\textbf{E}_4}{\partial t^2} \nonumber\\
&=& \mu_0\left(\frac{\partial}{\partial t}\nabla \otimes \textbf{M}_{\omega_4} + \frac{\partial^2 \textbf{P}_{\omega_4}}{\partial t^2} - c^2\nabla\left( \nabla \cdot \textbf{P}_{\omega_4} \right)\right].
\end{eqnarray}
which gives in the paraxial formulation, with beam 4 propagating in the $z$ direction, the following growth of the amplitude $E_4$:
\begin{eqnarray}
&&\left(\frac{\partial E_4}{\partial z} + \frac{1}{c}\frac{\partial E_4}{\partial t}\right) {\bf{u_4}} \nonumber\\
&=&
-\frac{i \mu_0 \omega_4}{2}\left[\left(c P_{\omega_4,x} + M_{\omega_4,y}\right) {\bf{u_x}} +\left(c P_{\omega_4,y} - M_{\omega_4,x}\right) {\bf{u_y}} \right]. \nonumber\\
\end{eqnarray}
The $x$ and $y$ subscripts stand for the $x$ and $y$ component. The same type of growth is obtained in four wave mixing in a standard medium where an effective susceptibility $\chi_v^3$
is defined and where we get:
\begin{equation}
\left(\frac{\partial E_4}{\partial z} + \frac{1}{c}\frac{\partial E_4}{\partial t}\right) {\bf{u_4}} = -\frac{i \omega_4}{2c} \chi_v^3 E_1 E_2 \overline{E}_3 {\bf{u_4}}.\label{Eq:E4_growth_chi3}
\end{equation}
The vacuum effective susceptibility thus corresponds to:
\begin{eqnarray}
\chi_v^3 &=& \frac{c \mu_0}{E_1 E_2 \overline{E_3}} \sqrt{\left(c P_{\omega_4,x} + M_{\omega_4,y}\right)^2 +\left(c P_{\omega_4,y} - M_{\omega_4,x}\right)^2},\nonumber\\
&=& \epsilon_0\sqrt{\left(K_{P,x} + K_{M,y} \right)^2 + \left(K_{P,y} - K_{M,x} \right)^2}
\end{eqnarray}
It depends on the $c_{i,j}$ parameters through the $\textbf{P}$ and $\textbf{M}$ vectors given in equations\,(\ref{Eq:P_omega4}) and (\ref{Eq:M_omega4}), or the $\textbf{K}_P$ and $\textbf{K}_M$ vectors given in equations\,(\ref{Eq:KP}) and (\ref{Eq:KM}). The scattered photon polarization is given by:
\begin{equation}
{\bf{u_4}} =
\frac{\left(c P_{\omega_4,x} + M_{\omega_4,y}\right) {\bf{u_x}} +\left(c P_{\omega_4,y} - M_{\omega_4,x}\right) {\bf{u_y}}} {\sqrt{\left(c P_{\omega_4,x} + M_{\omega_4,y}\right)^2 +\left(c P_{\omega_4,y} - M_{\omega_4,x}\right)^2}}. \label{Eq:Polar_u4}
\end{equation}
It also depends on the $c_{i,j}$ parameters.

Finally, the expected number of scattered photons is obtained by integrating equation\,(\ref{Eq:E4_growth_chi3}).
The result depends on the beams' profile (plane wave, gaussian beam,...), but it is always proportional to the square
of $\chi_v^3$ and proportional to the total cross-section of the process.

Experimentally, the choice of the laser setup and geometry is important to maximise the number of scattered photons and
to maximize the signal to noise ratio. But, to see more clearly the link between the $c_{i,j}$ coefficients and the number
of scattered photons, let's take some simple configurations with beam 2 and 3 counterpropagating with respect to beam 1.

If ${\bf{u_1}}={\bf{u_2}}={\bf{u_3}}$ and ${\bf{v_1}}=-{\bf{v_2}}=-{\bf{v_3}}$, one gets $\textbf{K}_P = 8c_{2,0}{\bf{u_1}} - 2 c_{1,1} {\bf{v_1}}$ and $\textbf{K}_M = 8c_{2,0}{\bf{v_1}} - 2 c_{1,1} {\bf{u_1}}$. The effective susceptibility is then:
\begin{equation}
\chi_{v,\mathrm{first}}^3 = 16\epsilon_0 c_{2,0}.\label{Eq:Chi3_first}
\end{equation}
The $c_{1,1}$ parameter cancelled out and $\chi_v^3$ only depends on $c_{2,0}$. A measurement in this configuration thus
allows to constrain this parameter independently from the others.

If ${\bf{u_1}}=-{\bf{v_2}}=-{\bf{v_3}}$ and ${\bf{v_1}}=-{\bf{u_2}}=-{\bf{u_3}}$, we get $\textbf{K}_P = 2c_{0,2}{\bf{u_1}} + 2 c_{1,1} {\bf{v_1}}$ and $\textbf{K}_M = 2c_{0,2}{\bf{v_1}} + 2 c_{1,1} {\bf{u_1}}$. The effective susceptibility is then:
\begin{equation}
\chi_{v,\mathrm{second}}^3 = 4\epsilon_0 c_{0,2}.\label{Eq:Chi3_second}
\end{equation}
It only depends on $c_{0,2}$.

Finally, if ${\bf{u_1}}={\bf{v_2}}={\bf{u_3}}$ and ${\bf{v_1}}={\bf{u_2}}=-{\bf{v_3}}$, we get $\textbf{K}_P = (4c_{2,0}-c_{0,2}){\bf{v_1}} + 2 c_{1,1} {\bf{u_1}}$ and $\textbf{K}_M = -(4c_{2,0}-c_{0,2}){\bf{u_1}} - 2 c_{1,1} {\bf{v_1}}$. The effective susceptibility is then:
\begin{equation}
\chi_{v,\mathrm{third}}^3 = \sqrt{2}\epsilon_0 (4c_{2,0}-c_{0,2}).\label{Eq:Chi3_third}
\end{equation}
It now depends on a linear combination of $c_{2,0}$ and $c_{0,2}$.

For more complicated laser beam configurations, the number of scattered photons $N_{\gamma\gamma}$ is of the form:
\begin{eqnarray}
N_{\gamma\gamma} &\propto& \left(\chi_{v}^3\right)^2,\\
&\propto& a c_{2,0}^2 + b c_{0,2}^2 + c c_{1,1}^2 \nonumber\\ &&+2dc_{2,0}c_{0,2} +2ec_{0,2}c_{1,1} +2fc_{2,0}c_{1,1}. \label{Eq:N_gg}
\end{eqnarray}
The $c_{0,1}$ parameter is absent. No limit or measurement on this coefficient can thus be given by photon-photon scattering experiments. In principle, studying the scattered photon polarization, given by equation\,(\ref{Eq:Polar_u4}), would allow to extract further informations on the different parameters $c_{2,0}$, $c_{0,2}$ and $c_{1,1}$.

The best experimental limit is reported in 2000\,\cite{Bernard_2000}. The value is compatible with zero. The error is about 18 orders of magnitude higher than the prediction of the QED prediction which corresponds to $c_{2,0}$ and $c_{0,2}$ given in equations\,(\ref{Eq:c20_HE_number}) and (\ref{Eq:c02_HE_number}), and $c_{1,1} = 0$.

\subsection{Magnetic birefringence versus photon-photon scattering}

Among experiments on light propagation in vacuum,
the most sensitive one concerns the measurement of magnetic birefringence using polarimetry. While the others are at more than 14 orders of magnitude from the QED (Heisenberg-Euler) prediction
(14 orders of magnitude for the magnetic birefringence using separated arms interferometer,
18 orders of magnitude for photon-photon scattering cross-section), the measurement of the Cotton-Mouton effect
is at less than 2 orders of magnitude from the QED prediction.

One could then envisage to use the most sensitive measurement to put a constraint on the others, and more particularly on the photon-photon scattering cross-section. As said before, the measurement of the vacuum magnetic birefringence cannot by itself constrain separately $c_{0,2}$ and $c_{2,0}$. On the other hand, we have shown on
simple examples that the $\chi_v^3$ dependance on the $c_{i,j}$ coefficients depends on the laser beam configuration. Limits on vacuum magnetic birefringence cannot therefore be translated into limits on photon-photon scattering since the dependence of
the effects from the NLED free parameters are generally different. However, photon-photon scattering limits can be represented as exclusion regions, as done in figure\,\ref{Fig:c02_c20_Exp_limits} for vacuum magnetic birefringence measurements, closing further the allowed range in the parameter space. Experiments whose goal is to measure the vacuum magnetic birefringence or the photon-photon scattering cross, far from being redundant, are complementary to test NLED theories.

This point, although apparently simple, is not always
fully understood. As a matter of fact, the authors of references \cite{Bregant_2008, DellaValle_2014}
 declare that a measurement of vacuum magnetic birefringence can constrain the Heisenberg-Euler lagrangian parameters and consequently
photon-photon scattering cross-section which it is not correct, as we just explained.


\section{Pointlike particles}\label{Particle}

For the moment, the experiments devoted to the study of light propagation in vacuum have not been able to test the
Heisenberg-Euler lagrangian. However, experiments on vacuum magnetic birefringence are only at two orders of magnitude
from the QED prediction, and one can hope that they will be gained in the near future. Does it mean that the Heisenberg-Euler
lagrangian has not yet been tested ? It is admitted that QED is widely and successfully tested on bound systems, for example in
the hydrogen atom, and on isolated charged particles with for example the measurement of the anomalous magnetic dipole moment of
the electron. Does it correspond to a test of the Heisenberg-Euler lagrangian ?
Is there any space still open for alternative NLED theories ?

\subsection{General expressions}

In the presence of external electric and magnetic fields, the vacuum reacts. It becomes polarized and magnetized and thus modifies the electric and magnetic fields.
Let's first calculate the $\textbf{P}$ and $\textbf{M}$ vectors induced by a point-like particle of charge $Q$ and magnetic moment
$\boldsymbol{\mu}=\mu\mathbf{e}_z$. The corresponding external electric and magnetic fields are:
\begin{eqnarray}
\mathbf{E} &=& \frac{Q}{4\pi \epsilon_0 r^2}\mathbf{e}_r,\\
\mathbf{B} &=& \frac{\mu_0 \mu}{4\pi r^3} \left[3\left(\mathbf{e}_z .
\mathbf{e}_r\right)\mathbf{e}_r - \mathbf{e}_z\right]\\
&=& \frac{\mu_0 \mu}{4\pi r^3} \left(3\cos \theta\mathbf{e}_r - \mathbf{e}_z\right)
\end{eqnarray}

To keep the validity of our non-linear lagrangian development, we only consider an electric field and a magnetic field well below the critical ones defined in the Heisenberg-Euler lagrangian.
We therefore assume that $r\gg r_\mathrm{cr}^E$ and $r\gg r_\mathrm{cr}^B$ with $r_\mathrm{cr}^E = \sqrt{Q/4\pi \epsilon_0 E_\mathrm{cr}}$
and $r_\mathrm{cr}^B = \left(\mu_0\mu/4\pi B_\mathrm{cr}\right)^{1/3}$. For a proton, $Q = 1.6\times 10^{-19}$\,C
and $\mu = 1.41 \times 10^{-26}$\,J.T$^{-1}$, and one obtains $r_{cr}^E \sim 3\times10^{-14}$\,m
and $r_{cr}^B \sim 7\times10^{-15}$\,m.

Injecting the previous electric and magnetic fields in the Lorentz invariants given by equations\,(\ref{Eq:F}) and (\ref{Eq:G}), we get:
\begin{eqnarray}
\mathcal{F} &=& \frac{Q^2}{(4\pi)^2\epsilon_0 r^4}\left[1-\left(\frac{\mu}{cQr} \right)^2 \left(1+3 \cos^2\theta \right) \right]\\
\mathcal{G} &=& \sqrt{\frac{\mu_0}{\epsilon_0}} \frac{Q 2\mu  \cos\theta}{(4\pi)^2 r^5}.
\end{eqnarray}
The corresponding $\textbf{P}$ and $\textbf{M}$ vectors are:
\begin{eqnarray}\label{Eq:P_Proton}
\textbf{P} &=& c_{0,1} \sqrt{\epsilon_0 \mu_0}\frac{\mu}{4\pi r^3}  \left(3\cos \theta\mathbf{e}_r - \mathbf{e}_z\right) \nonumber\\
&+& c_{2,0} \epsilon_0 \textbf{E} \frac{Q^2}{4\pi^2\epsilon_0 r^4}\left[1-\left(\frac{\mu}{cQr} \right)^2 (1+3\cos^2 \theta) \right] \nonumber\\
&+& c_{0,2} \epsilon_0 E \frac{\mu_0\mu^2 \cos \theta}{4\pi^2 r^6}(3 \cos \theta \textbf{e}_r - \textbf{e}_z) \nonumber\\
&+& c_{1,1} \epsilon_0 \textbf{E} \sqrt{\frac{\mu_0}{\epsilon_0}}\frac{Q\mu \cos \theta}{4\pi^2 r^5} \nonumber\\
&+& c_{1,1} \epsilon_0 E \sqrt{\frac{\epsilon_0}{\mu_0}} \frac{Q\mu_0\mu}{(4\pi)^2\epsilon_0 r^5} \left[1-\left(\frac{\mu}{cQr} \right)^2 (1+3\cos^2 \theta) \right] \nonumber\\
&&(3 \cos \theta \textbf{e}_r - \textbf{e}_z) \nonumber\\
\end{eqnarray}
\begin{eqnarray}\label{Eq:M_Proton}
\textbf{M} & = & c_{0,1} \sqrt{\frac{\epsilon_0}{\mu_0}}\frac{Q}{4\pi \epsilon_0 r^2} \textbf{e}_r \nonumber\\
&-& c_{2,0} \frac{\textbf{B}}{\mu_0} \frac{Q^2}{4\pi^2\epsilon_0 r^4}\left[1-\left(\frac{\mu}{cQr} \right)^2 (1+3\cos^2 \theta) \right] \nonumber\\
&+& c_{0,2} \frac{B(\theta = 0)}{\mu_0} \frac{Q^2 \cos \theta}{8 \pi^2 \epsilon_0 r^4} \textbf{e}_r \nonumber\\
&-& c_{1,1} \frac{\textbf{B}}{\mu_0} \sqrt{\frac{\mu_0}{\epsilon_0}}\frac{Q\mu \cos \theta}{4\pi^2 r^5} \nonumber\\
&+& c_{1,1} \frac{B(\theta = 0)}{\mu_0} \sqrt{\frac{\epsilon_0}{\mu_0}} \frac{Q^3}{32\pi^2\epsilon_0^2 \mu r^3} \nonumber\\
&&\left[1-\left(\frac{\mu}{cQr} \right)^2 (1+3\cos^2 \theta) \right] \textbf{e}_r, \nonumber\\
\end{eqnarray}
with $B(\theta=0) = \mu_0\mu/2\pi r^3$.

The electric and magnetic fields are slightly modified by the polarization and magnetization of the vacuum and become:
\begin{eqnarray}
\mathbf{E_\mathrm{V}} &=& \mathbf{E} - \frac{\textbf{P}}{\epsilon_0},\\
\mathbf{B_\mathrm{V}} &=& \mathbf{B} + \mu_0 \textbf{M}.
\end{eqnarray}
Some of the corrections to the fields given in the previous equations have a form that is very unusual,
like for example the radial correction to $\textbf{M}$. These unusual corrections are related to $(\textbf{E} \cdot \textbf{B})$ and $c_{0,2}$.

\subsection{Electric dipole moment and magnetic monopole}

We first focus on the first term of equations\,(\ref{Eq:P_Proton}) and (\ref{Eq:M_Proton}) proportional to the $c_{0,1}$ coefficient:
\begin{eqnarray}
\textbf{P}_{01} &=& c_{0,1} \sqrt{\epsilon_0 \mu_0}\frac{\mu}{4\pi r^3}  \left(3\cos \theta\mathbf{e}_r - \mathbf{e}_z\right)\\
&=& c_{0,1}\sqrt{\frac{\epsilon_0}{\mu_0}}\textbf{B},\\
\textbf{M}_{01} & = & c_{0,1} \sqrt{\frac{\epsilon_0}{\mu_0}}\frac{Q}{4\pi \epsilon_0 r^2}\textbf{e}_r\\
&=& c_{0,1}\sqrt{\frac{\epsilon_0}{\mu_0}}\textbf{E}.
\end{eqnarray}
If $c_{0,1}$ is not zero, as soon as an electric field {\bf E} and a magnetic field {\bf B} are superimposed in a vacuum, a non linear term appears inducing a
correction to {\bf E} proportional to {\bf B} and a correction to {\bf B} proportional to {\bf E}.
So, for the case of an isolated particle of charge $Q$ and magnetic moment $\mu$, if the $c_{0,1}$ parameter is not zero, the magnetic
dipole field should also appear as an electric dipole field so that the particle acquires an electric dipolar moment:
\begin{equation}
\textbf{d} = \frac{c_{0,1}}{c} \boldsymbol{\mu}.
\end{equation}
On the other hand, the radial electric field should induce a radial magnetic field so that the particle acquires a magnetic monopole:
\begin{equation}
m = c_{0,1} Q c,
\end{equation}
where we write the monopole radial field $\textbf{B}_\mathrm{m}$ as $\textbf{B}_\mathrm{m} = \mu_0 m/ 4\pi r^2 \textbf{e}_r$.

The standard model predicts a non-zero electric dipole moment for the electron, muon or tau particles, due to $CP$ violation.
The predicted value is however well below the current experimental sensitivities. For example, for the electron one expects $d_e \simeq 10^{-38}$\,e\,cm \cite{Pospelov_2005}.
As far as we understand, a $c_{0,1} \simeq 10^{-28} $ would therefore mimic the standard model EDM for the electron. No experiment has ever detected this deviation, but constraints can be found. Some of them are listed in Table\,\ref{Tab:EDM}
with the corresponding limit on $c_{0,1}$ (see also the particle data book \cite{Olive_2014}).

\begin{center}
\begin{table}[h]
\begin{center}
\begin{tabular}{m{1.3cm} m{3cm} m{1cm} m{3cm}}
\hline \hline
\tabularnewline

Particle &  $d$ (e\,cm) & Ref. & $c_{0,1}$\\

\hline
\tabularnewline
electron & $< 10.5\times 10^{-28}$ & \cite{Hudson_2011} & $< 5.43\times 10^{-17}$\\

\tabularnewline
muon & $(-0.1 \pm 0.9) \times 10^{-19}$ & \cite{Bennett_2009} & $(1.1 \pm 9.6) \times 10^{-7}$\\

\tabularnewline  tau & $-0.22$ to $0.45 \times 10^{-16}$ & \cite{Inami_2003} & $-8.1$ to $4 \times 10^{-3}$\\

\tabularnewline  proton & $< 7.9 \times 10^{-25}$ & \cite{Griffith_2009} & $< 2.69 \times 10^{-11}$\\

\tabularnewline \hline
\hline
\end{tabular}
\end{center}
\caption{Constraints on electric dipole moment of charged particles and corresponding constraints on the $c_{0,1}$ coefficient.}
\label{Tab:EDM}
\end{table}
\end{center}

Concerning magnetic monopoles, they have been first introduced by P.\,A.\,M. Dirac in 1931 \cite{Dirac_1931}.
The goal was to explain the quantization of electric charge by postulating the existence of an elementary magnetic charge,
 $Q_\mathrm{M}^\mathrm{D} = 2\pi \hbar/e$, that is now called the Dirac charge. More recently,
it was understood that
in the framework of Grand Unification Theories (GUT) the electric and magnetic charges are naturally quantized \cite{Hooft_1974}.

From an experimental point of view, limits exist for electron and proton magnetic charge\,\cite{Lorin_1968, Olive_2014}. The electron magnetic
charge $Q_\mathrm{M}$, inducing a Coulomb magnetic field $\textbf{B} = Q_M/4\pi r^2 \textbf{e}_r$, has been found to be:
\begin{equation}
Q_\mathrm{M} < 4\times10^{
-24}
 Q_\mathrm{M}^\mathrm{D}.
\end{equation}
This corresponds to:
\begin{equation}
c_{0,1} < 3 \times 10^{-22}.
\end{equation}
which is a stronger limit than the one obtained by EDM search.

\subsection{Bound system and Lamb shift}

For the sake of simplicity,
we now consider $c_{0,1}$ and $c_{1,1}$ to be zero, or at least negligible. Using equations\,(\ref{Eq:P_Proton}) and (\ref{Eq:M_Proton}), the $\mathbf{E_\mathrm{V}}$ and $\mathbf{B_\mathrm{V}}$ vectors can be approximated,
at the leading order, to:
\begin{eqnarray}
\mathbf{E_\mathrm{V}} &=& \textbf{E} \left[1 - c_{2,0} \frac{Q^2}{4\pi^2\epsilon_0 r^4}\right] \label{Eq:D_Proton}\\
\mathbf{B_\mathrm{V}} & = & \textbf{B} \left[ 1
- c_{2,0} \frac{Q^2}{4\pi^2\epsilon_0 r^4}\right]
+
c_{0,2} \frac{B(\theta = 0)}{\mu_0} \frac{Q^2 \cos \theta}{8 \pi^2 \epsilon_0 r^4} \textbf{e}_r
 \nonumber\\
\end{eqnarray}

Let's first discuss the implications of equation\,(\ref{Eq:D_Proton}). The correction in the Coulomb potential energy is proportional to $1/r^5$:
\begin{equation}
\delta V = -c_{2,0}\frac{Q^3}{80 \pi^3 \epsilon_0^2 r^5}.
\end{equation}
In the QED framework, one obtains:
\begin{equation}
\delta V_\mathrm{QED} = -\frac{Q}{4\pi \epsilon_0 r}\frac{2 \alpha^3}{225 \pi} \left(\frac{\hbar}{m_e c r}\right)^4.
\end{equation}
This correction has been studied since 1956\,\cite{Wichmann_1956} and it is called the Wichmann-Kroll potential.

This correction, proportional to $c_{2,0}$, induces
an energy shift in bound systems and it is indeed part of the well-known Lamb shift. In Table\,\ref{Tab:LambShift}, we give some examples of the contribution
of the Wichmann-Kroll correction to the leading term for different energy transitions and different systems. We also add the corresponding experimental precision.

\begin{center}
\begin{table}[h]
\begin{center}
\begin{tabular}{m{1.8cm} m{2.4cm} m{2cm} m{2cm}}
\hline \hline
\tabularnewline

System and &  Wichmann-Kroll & Experimental & Remarks\\
energy levels&  contribution to & relative & \\
 &  the leading term & uncertainty & \\

\hline
\tabularnewline
H & 0.3\,ppm & 3\,ppm \cite{Biraben_2009} & \\
1S & & & \\
\tabularnewline
H muonic & 5 ppm &  15 ppm \cite{Pohl_2010} & Proton charge\\
2S-2P & & & radius puzzle\\
\tabularnewline \hline
\hline
\end{tabular}
\end{center}
\caption{Examples of the contribution of the Wichmann-Kroll correction to the Lamb shift leading term for two different energy transitions and systems, to be compared to the relative uncertainties obtained on the Lamb shift measurements\,\cite{Julien_2012}.}
\label{Tab:LambShift}
\end{table}
\end{center}

In the case of the Lamb shift of the 1S and 2S level in atomic hydrogen, the Wichmann-Kroll correction has been calculated to be 0.3\,ppm of the
leading term \cite{Mohr_2012,Julien_2012}, while the corresponding measurements have a precision of about 3\,ppm\,\cite{Biraben_2009,Julien_2012}. All these calculations has been performed
in the accepted QED framework with $c_{2,0}$ given by equation\,(\ref{Eq:c20_HE_number}). It is worth stressing that $c_{0,2}$ cannot be constrained by bound systems studies.

\begin{figure}[h]
\begin{center}
\includegraphics[width=8cm]{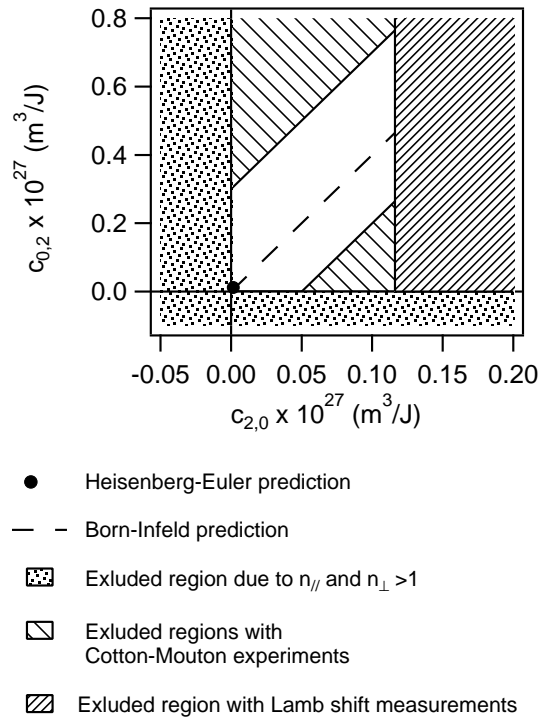}
\caption{\label{Fig:c02_c20_Lamb_shift} Best experimental limits on $c_{0,2}$ and $c_{2,0}$ parameters. The excluded region due to Lamb shift measurements is added.}
\end{center}
\end{figure}

Now the $c_{2,0}$ dependence of the Wichmann-Kroll correction to the Lamb shift is linear \cite{Wichmann_1956}. This
means that the measurement of the 1S-2S Lamb-shift in hydrogen, presented in Table \ref{Tab:LambShift}, constraints the value as follows: $c_{2,0} < 10\times c_{2,0}^\mathrm{HE}$.
We add the corresponding excluded region in figure\,\ref{Fig:c02_c20_Lamb_shift}.

In the case of the 2S-2P lamb shift of muonic hydrogen, the correction is evaluated at 5 ppm \cite{Eides_2001} but the measurement is at 15 ppm \cite{Pohl_2010}. Furthermore the proton radius extracted from this measurement is not in agreement with the one inferred from the hydrogen measurement. This is an important issue that is now called the ``proton charge radius puzzle''.
This means that Wichmann-Kroll correction has not been tested and therefore there are not yet further informations on $c_{2,0}$ coming from QED tests in bound systems.

Let's come to the modification to the magnetic field.
It looks like nobody as ever consider it except Jeremy Heyl as modification of a macroscopic magnetic dipole \cite{Heyl_1997} but without the term proportional to $c_{0,2}$ coming from the coupling between the electric and the magnetic field. This term has a very unusual form. No calculation of
energy shift induced by this correction exists, although this modification of the magnetic field of a pointlike charge should affect at least the atomic hyperfine splitting. In fact, the leading term in this energy splitting, called the Fermi term \cite{Fermi_1930}, is proportional to the field due to the bound particle at the position of the nucleus. In the case of the hydrogen atom, the correction of the electron magnetic field at a distance of a
Bohr radius is of the order of $2\times10^{-17}$, when the precision of the hydrogen ground state hyperfine splitting measurement is of the order of 10$^{-13}$ \cite{Essen_1971}. For the muonic hydrogen the correction of the muon magnetic
field at the position of the proton is of the order of $4\times 10^{-8}$ but the ground state hyperfine splitting of muonic hydrogen has not yet been
measured (see e.g. \cite{Bakalov_2015}).

%

\subsection{Limits on the Born-Infeld $E_{abs}$ free parameters}

The Born Infeld NLED is constructed on the assumption that an absolute electric field exists in nature. Atomic energy levels should therefore be
different from the ones predicted without such a field limitation. The natural way to constrain such a free parameter is therefore to look
for the predicted energy variation in high atomic number atoms where non linearities should be more important. This has been done in 1973 by Soff,
Rafelski and Greiner \cite{Soff_1973} who report that $E_\mathrm{abs}$ has to be greater than
$1.7\times10^{22}$\,V/m. More recently their results have been questioned \cite{Carley_2006} even if the authors agree that the value proposed by Born and Infeld is not physically viable. For the sake of argument, let's note that an $E_\mathrm{abs} = 1.7\times10^{22} $\,V/m corresponds to a $c_{2,0}$ of about 5 orders of magnitude smaller than
the one predicted by QED.

\section{Conclusion}

In this paper we set a framework in which experiments whose goal is to test QED predictions can be used in a more general way to test NLED
which contains low-energy QED as a special case. We review some of these experiments and we establish limits on the different free parameters
$c_{0,1}$, $c_{2,0}$, $c_{0,2}$ and $c_{1,1}$, generalizing QED predictions in the framework of NLED.
Actually only $c_{0,1}$, $c_{2,0}$ and $c_{0,2}$  can be constrained. As far as we know, no experiment constraining $c_{1,1}$ exists.


The parametrization of the photon-photon interaction lagrangian is also very useful to understand the mutual impact of QED tests of different nature.
In particular, we show that $c_{2,0}$ can be limited by measurements of Wichmann-Kroll potential corrections, as in the case of the
1S-2S Lamb-shift in atomic hydrogen,
at a level that is compatible with limits coming from vacuum magnetic birefringence.

The Heisenberg-Euler lagrangian is a special case of NLED. In bound systems it is related to the Wichmann-Kroll potential that is the correction to the Coulomb potential at large distances.
The leading term to the Coulomb potential corrections is given by the Uehling potential representing the short distance corrections.
The Wichmann-Kroll potential induces lower orders corrections than the Uehling ones and that is why, while in general one can say that QED in bound
systems is well tested, this is not true specifically for the long distance corrections where the direct tests of NLED come into play.
Of course, the Wichmann-Kroll potential and the Uehling one come both from the same theoretical framework and it is difficult to imagine that the short
range regime is well treated while the long range is not, but nevertheless one has to test if some new physics appears in the long range inducing
corrections not predicted by standard QED. This looks as an important task largely justifying NLED direct tests.

Let's finish with the anomalous magnetic moment $(g-2)$ of isolated particles which is one of the best tested quantity in QED\,\cite{Hanneke_2008}. As discussed, for example, in reference\,\cite{Schlenvoigt_2016}, photon-photon scattering contributes as a subdiagram to the $g-2$ and the Lamb shift. At first sight, it thus seems feasible to use $g-2$ measurements to constrain the $c_{0,2}$ or $c_{2,0}$ parameters, as done in the previous section with the Lamb shift.
However, the $g-2$ of isolated particles corresponds to a physical quantity that is related to short range physics, as far as we understand, or at least the long range corrections have never been stated explicitly as in the case
of the Wichmann-Kroll corrections for the Lamb shift in bound systems.
Furthermore, the $g-2$ corrections change
the value of the magnetic moment, not the shape of the dipolar field. But the correction given in equation\,(\ref{Eq:M_Proton})
indicates a change in the shape of the field.
It is not clear to us how to combine both of them. This is certainly an important point to clarify.

\acknowledgments

We thank Jeremy Heyl for very useful discussions.
We acknowledge the support of the \textit{Fondation pour la recherche IXCORE} and the ANR (Grant No. ANR-14-CE32-0006).

\bibliography{biblio}

\end{document}